\documentstyle[11pt,newpasp,twoside,epsf,psfig]{article}
\def\edcomment#1{\iffalse\marginpar{\raggedright\sl#1\/}\else\relax\fi}
\reversemarginpar

\begin{document}
\title{The Parameter Space Occupied by Galaxies: The optical window}
\author{Simon Driver}
\affil{School of Physics and Astronomy, North Haugh, University of St Andrews,
Fife, KY16 9SS, SCOTLAND}
\author{Nicholas Cross}
\affil{School of Physics and Astronomy, North Haugh, University of St Andrews,
Fife, KY16 9SS, SCOTLAND}

\begin{abstract}
We discuss the optical parameter space (luminosity and surface brightness) over
which galaxies are known to exist and report on the results from three recent
datasets: the Two-degree Field Galaxy Redshift Survey, the Hubble Deep Field
north, and the Local Group. These data are combined to provide a unique and
comprehensive insight into the luminosity distribution of galaxies over
the range $-24 < M_{B} < -8$ mags,  and the surface brightness distribution 
over
the range $18 < \mu_{e} < 28$ mags$/\Box''$. The main conclusions are:
(1) luminous low surface brightness galaxies are rare;
(2) there exists a universal luminosity-surface brightness relation;
(3) the most numerous galaxies are low luminosity low surface brightness 
dwarfs.
\end{abstract}

\section{Introduction}
The study of galaxy formation and evolution cannot commence in earnest until 
we have a full and complete description of the local galaxy population - our 
benchmark if you will. Until we've fully defined the local ``zoo of galaxies'' 
and their relative abundances we can never be sure that any trend in redshift
cannot be interpreted as ``selection bias'' rather than ``evolution''.
So how close are we getting to a complete description ?
Typically the galaxy population is quantified via the following 
observable parameters: luminosity, surface brightness, morphology and colour. 
These are in turn loosely associated with the fundamental properties of: mass, 
angular momentum, evolutionary history and age, respectively. In this article 
we review our current insight into the parameter space occupied by galaxies
in terms of their luminosity and surface brightness distributions. We argue
that in order to unambiguously remove the spectre of selection bias, the
luminosity and surface brightness distributions must be considered 
simultaneously. This is borne out by three recent datasets: the Two-degree 
Field Galaxy Redshift Survey (2dFGRS), the Hubble Deep Field (HDF), and the 
latest Local Group (LG) census. Combining these datasets provides the current 
state-of-the art insight into the low redshift galaxy population.

\section{The current state-of-play}

\subsection{The Luminosity Distribution}
The number density of galaxies as a function of luminosity is conveniently 
described by a Schechter function (Schechter 1976), i.e.:

~

\begin{center}
$\phi(L)dL = \phi_* (\frac{L}{L_*})^\alpha e^{-(\frac{L}{L_*})} dL$,
\end{center}

~

\noindent
this effectively reduces the entire galaxy population to three crucial defining 
parameters: $\phi_*, L_*$ and $\alpha$. These represent the
density and luminosity calibration points, and the faint-end slope respectively.
The main asset of the Schechter function is its simplicity, and the ability to 
thereafter calculate the galaxian luminosity-density ($j$) and mass-density
($\rho$):

~

\begin{center}
$j = \phi_* L_* \Gamma(\alpha+2)$, \hspace{1.0cm} and \hspace{1.0cm} $\rho = (\frac{M}{L}) j$,
\end{center}

~

\noindent
The ease with which these fundamentally important cosmological parameters can 
be extracted from a galaxy catalogue is however beguiling. Table 1 and Figure 1 
show a compendium of recent measurements of these three key parameters and the 
implied luminosity- and mass- densities 
(expressed in terms of the critical density assuming $\frac{M}{L} \sim 250$).
The variation in these parameters is a cause for alarm.

\begin{table}[h]
\caption{Summary of key values derived from redshift surveys}
\begin{center}

\begin{tabular}{llccccc} \tableline
Survey & Reference & $M_{*}$ & $\phi_{*}$ & $\alpha$ & $j_{B}$ & $\Omega_{M}$ \\ 
       &           & (mags)  & (Mpc$^{-3}$) &        & ($10^8hL_\odot$Mpc$^{-3}$) & \\ \tableline
SSRS2   & Marzke et al. (1998)   & -19.43 & 0.013 & -1.12 & 1.28 & 0.12 \\
UKST    & Ratcliffe et al (1998) & -19.63 & 0.017 & -1.04 & 2.02 & 0.19 \\ 
ESP     & Zucca et al (1997)     & -19.61 & 0.020 & -1.22 & 2.58 & 0.23 \\
LCRS*   & Lin et al (1996)       & -20.29 & 0.019 & -0.70 & 0.87 & 0.08 \\
EEP     & Efstathiou et al (1988)& -19.68 & 0.016 & -1.07 & 1.89 & 0.17 \\
APM     & Loveday et al (1995)   & -19.50 & 0.014 & -0.97 & 1.35 & 0.12 \\  
Afib    & Ellis et al (1996)     & -19.20 & 0.026 & -1.09 & 2.05 & 0.19 \\
CfA     & Marzke et al (1994)    & -18.8  & 0.040 & -1.00 & 2.06 & 0.19 \\ \tableline \tableline
\end{tabular}
\end{center}

\noindent
*: A mean $(b-r)=1.5$ has been assumed.
\end{table}

\noindent
So what's going on ? There are two prime candidates for the large
variations seen in Table 1:
(1) our galaxy catalogues are plagued by selection biases;
(2) clustering in the universe is much more severe than previously thought. 
The most lamented selection bias is that of surface brightness 
incompleteness (i.e. galaxies are either missing from the original imaging
survey, their magnitudes are severely underestimated or their redshifts were 
unobtainable). Similarly although 
clustering has been known since the pioneering work of de Lapparent, Geller 
\& Huchra (1986) no survey has so far been sufficiently large as to establish 
the scale at which large-scale structure averages out. Of these problems only
the latter will be solved by the new generation surveys currently underway 
(i.e the 2dFGRS and the SDSS).

At this point it is also worth noting that, accepting the discrepancies in 
the existing surveys, they also only probe over a narrow range of 
luminosity ($-21 < M_{B} < -17$). However galaxies are known with absolute 
magnitudes of $M_{B} = -6$ (Mateo 1998). Hence we have only charted, and 
poorly at that, one quarter of the luminosity range over which galaxies are 
known to exist. {\bf Our assessment of the local galaxy population is both 
inaccurate and incomplete.}

\begin{figure}
\plotfiddle{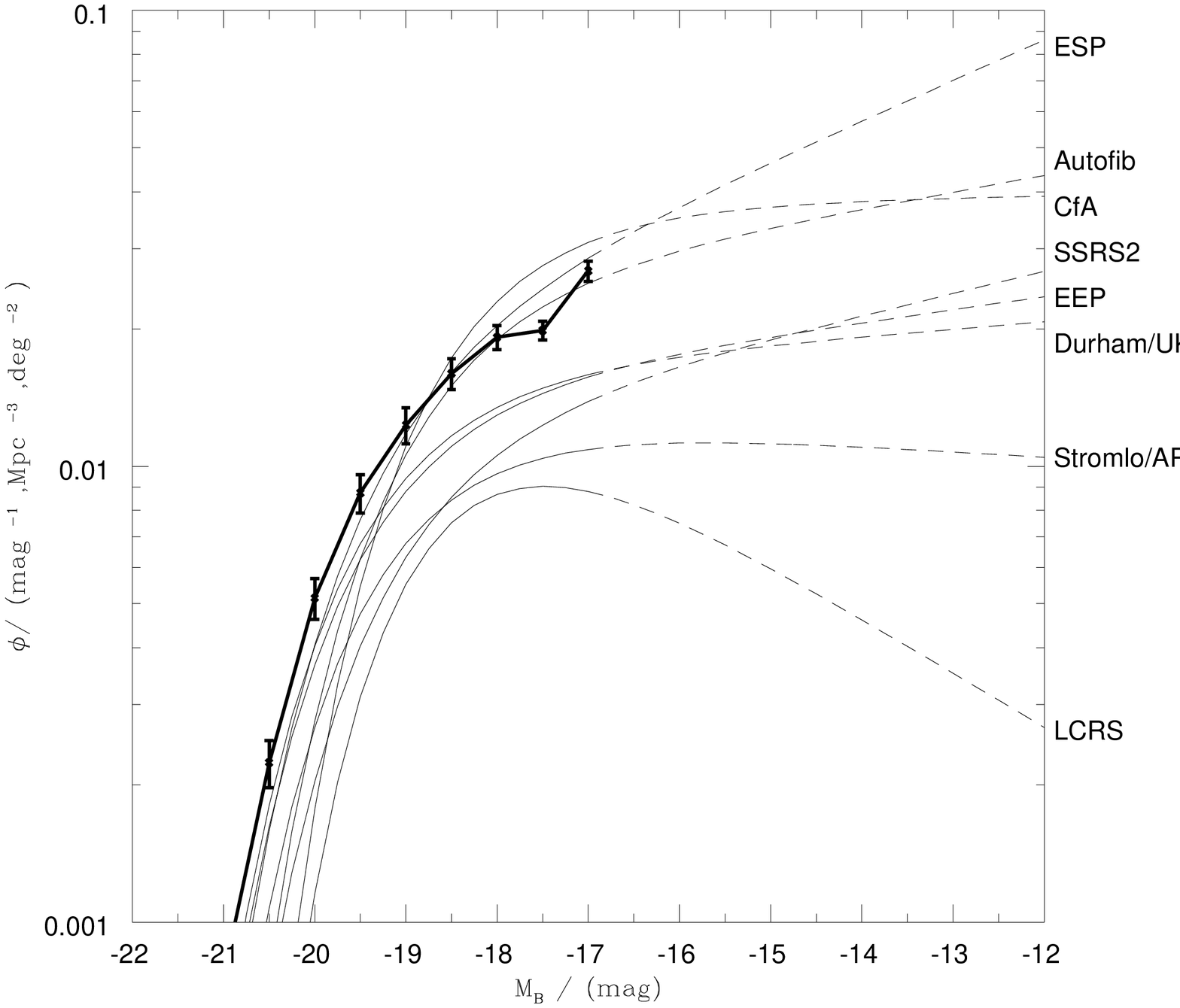}{100mm}{0}{85}{57}{-250}{-85}
\caption{A comparison of recent luminosity function measures (see Table 1 
for references). The solid line shows the luminosity range over which data 
was obtained and the dashed lines are extrapolations. The thick line shows 
the latest results from the 2dFGRS corrected for surface brightness selection 
effects and clustering}
\plotfiddle{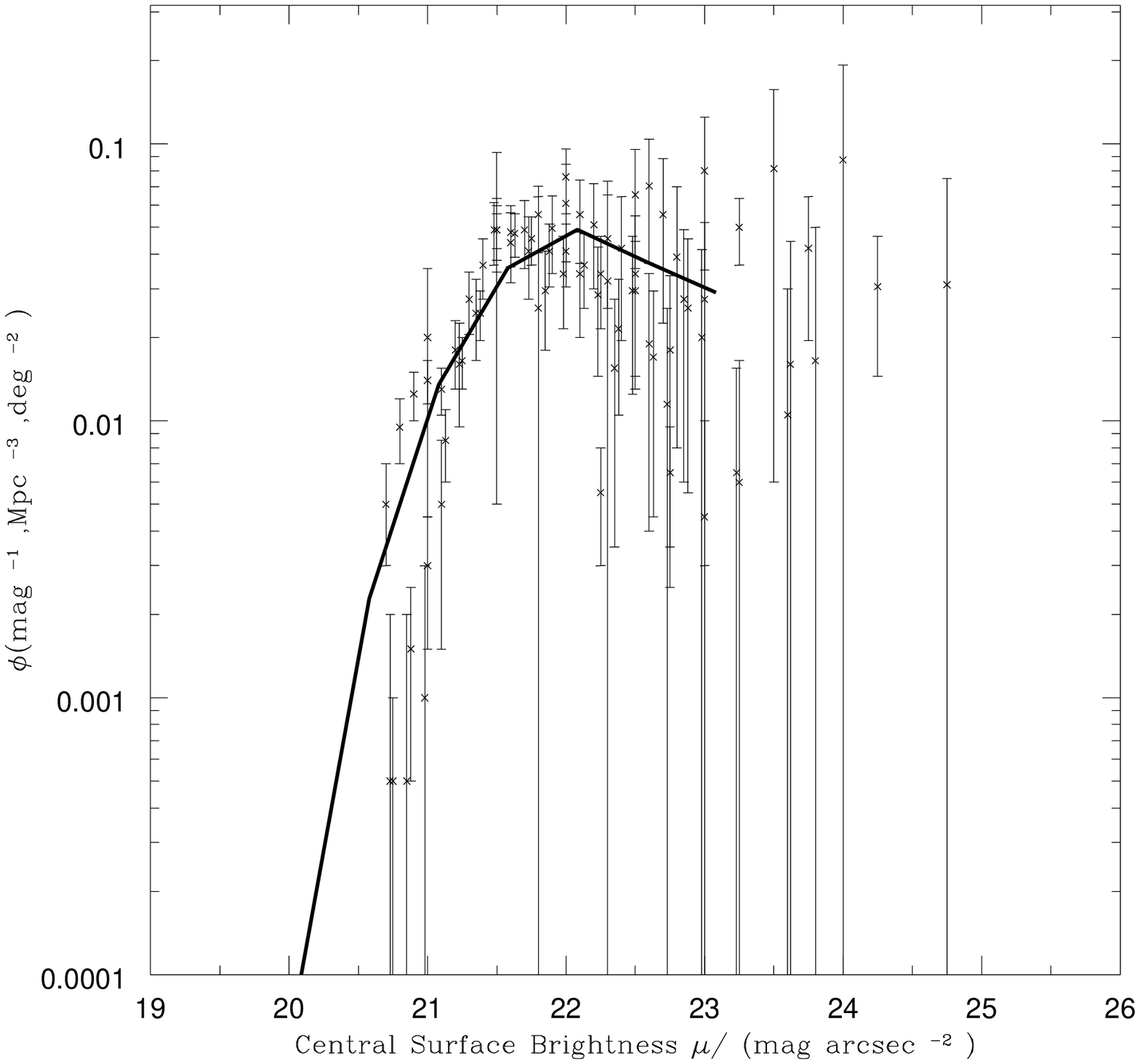}{100mm}{0}{80}{52}{-250}{-85}
\caption{The surface brightness distribution of galaxies adapted from O'Neil \&
Bothun (2000). The solid line shows the latest result from the Two-degree 
Field Redshift Survey for galaxies with $M_{B} < -17$ mags (Cross et al 2000).}
\end{figure}

\subsection{The Surface Brightness Distribution}
The surface brightness distribution of galaxies is really a by-product or
outcome of the raging debate over the impact of surface brightness selection
biases on galaxy catalogues (as introduced in \S 2). The debate originates
from the observation by Ken Freeman that most spiral galaxies appeared to
have a central disk surface brightness close to 21.7 B mags$/\Box''$ (Freeman 
1970). This led to the conjecture (Disney 1976) and formulation
(Disney \& Phillipps 1987) that there exists an optimal surface brightness
window through which galaxies are readily observable. This selection window
was shown to be dependent on the isophotal depth of the survey and redshift
(for a full 
description of visibility theory see Phillipps, Davies \& Disney 1990, or the 
summary in Cross et al 2000). Outside of this narrow window galaxies are 
either too compact to be distinguished from stars or too diffuse to be 
detectable. More recently the cause has been taken up by others
(see review by Impey \& Bothun 1996) and recent surveys, which push to faint 
limiting isophotes, have indeed uncovered populations of low surface 
brightness galaxies - for example, Sprayberry et al (1996), Dalcanton et al 
(1997), and O'Neil \& Bothun (2000) and references therein.

One claim from the low surface brightness community is that a substantial
amount of luminosity and mass may reside in very luminous very low surface 
brightness galaxies, hitherto overlooked in the spectroscopic surveys listed 
in Table 1. This is indeed plausible as the source material for all local 
spectroscopic surveys are photographic plates scanned with bright isophotal 
limits ($\sim 24.5$ B mags/sq arcsec). This conjecture is fueled from the 
serendipitous discovery of Malin 1 (see Bothun et al 1987) - the most luminous 
and HI massive field galaxy known. This ``crouching giant'', was originally 
thought to be a dwarf elliptical in the Virgo cluster. Spectroscopic follow-up 
revealed that it lay well beyond the cluster and subsequent deep imaging 
revealed a spectacular low surface brightness spiral disk with a central 
surface brightness of $\sim 25.5$ V mags/sq arcsec and an unprecedented scale 
length of 55 kpcs. Malin 1 itself is easily detectable, because of its active 
AGN core, however its extensive disk is invisible to photographic surveys. 
This raises two concerns; firstly that Malin 1's without cores may exist,
and secondly that many of the systems we currently accept as dwarfs
might actually represent just the cores of other crouching giants.
The blind HI surveys are the most promising method for searching for these
systems and fortuitously the papers presented at this meeting have, in my 
mind, gone a long way towards demonstrating that core-less Malin 1 type 
objects are very 
rare (see articles in this proceedings by Webster, Schneider and Tully).

Figure 2 shows the latest compendium of the surface brightness distribution 
of galaxies (adapted from O'Neil \& Bothun 2000). This is directly analogous 
to Fig. 1 except central surface brightness rather than 
luminosity is plotted. This figure has moved on a lot since the initial 
Freeman result and it demonstrates that surveys with bright 
isophotal cutoffs will incompletely sample the galaxy population. This
problem is exacerbated as one probes towards higher redshift. The
combination of $(1+z)^4$-dimming and K-corrections combine to radically
shift the selection windows through which we sample the underlying galaxy 
population.
{\bf Low surface brightness galaxies are missing in large numbers from our 
local galaxy catalogues}.

\subsection{Galaxy Morphology}
The Hubble tuning fork and its revision to include Magellanic type irregulars
has been the bread-and-butter of galaxy morphology with the majority of
all nearby galaxies being classified onto this system. However in recent 
years there has been a quiet revolution with the discovery of many galaxies
which fall outside of the revised tuning fork (Ferguson \& Binggeli 1994).
As we have probed deeper 
into the local universe, numerous varieties of dwarf galaxies have emerged.
Figure 3 shows the absurdly confusing picture when the Hubble tuning fork
is further revised to include the full zoo of commonly used morphological 
nomenclature. Of further concern is the evidence from
the Hubble Deep Field which indicates that the Hubble sequence
is a recent phenomena representing only the luminous-minority of
possible types (Driver et al 1998). 
{\bf The Hubble tuning fork is an incomplete description
of the galaxy population}.

\begin{figure}[p]
\plotone{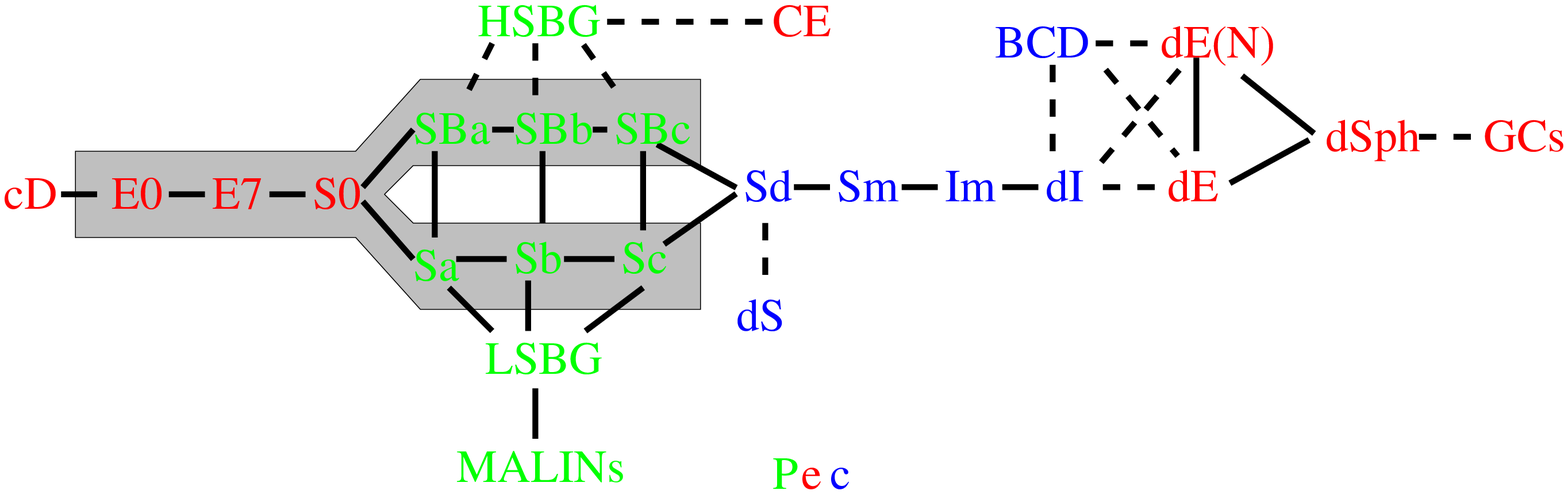}
\caption{The breakdown of the once elegant Hubble tuning fork system of galaxy
classification. Solid and dashed lines denoted suggested or speculative
evolutionary links respectively.}
\vspace{2.0cm}

\plotone{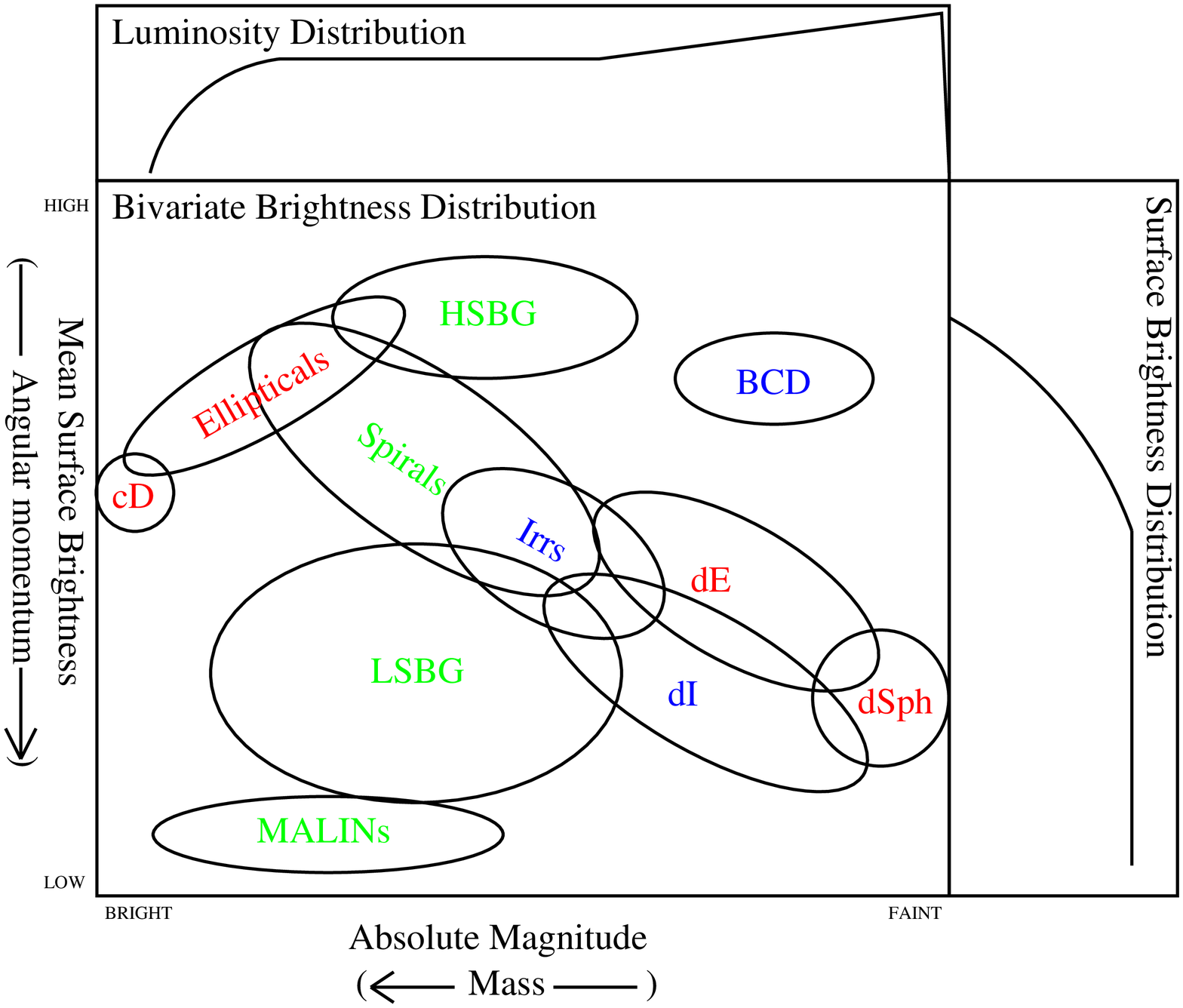}
\caption{The Galaxy Population Plot: a new way of looking at the 
galaxy population ?}
\end{figure}

\subsection{Recap'}
To recap the luminosity function of galaxies varies in amplitude at the
$L_{*}$ point by a factor of 2 with a faint end that flaps around from
$-0.8$ (Lin et al 1996) to $-1.25$ (Zucca et al 1997) and runs out of steam
at $M_{B} \sim -17$.
The surface brightness distribution of galaxies shows equal variation with
the champions of surface brightness arguing that we may be missing up to
50\% of the local galaxy population at all luminosities (although this is not
borne out by the HI surveys). Together these lead to an uncertainty in the 
local luminosity density of a factor of $\sim 6$. The impact on the 
mass-density is potentially worse given the high mass-to-light ratios of 
both low luminosity (Mateo 1998) and low surface brightness (Zwaan et al 1995; 
de Blok et al 1996) galaxies yielding a potential order of magnitude
uncertainty in $\Omega_{M}$ (as derived from existing luminosity functions, 
e.g. Carlberg  et al 1996; Persic \& Salluci 1992; Fukugita, Hogan \& Peebles 
1998). In addition the simple elegance of the original Hubble tuning-fork has 
been severely compromised by the discovery that the majority of galaxies, in 
terms of numbers, lie outside of the Hubble tuning fork (Driver 1999).

{\bf Given this degree of uncertainty the study of galaxy evolution via 
statistical methods (e.g. galaxy counts, redshift distributions and 
comparisons between high and low redshift samples) is fundamentally flawed.}

\section{Combining Luminosity, Surface Brightness and Morphology}
The conclusion from Figs 1 \& 2 is that galaxies occupy a range in both
luminosity and surface brightness. But the figures, as well as containing 
unsatisfactory scatter, do not indicate whether 
the surface brightness distribution of Figure 2 exists at all luminosity 
intervals (and vice versa). For example although Fig. 2 indicates that low 
surface brightness galaxies exist, do they really exist at bright 
luminosities ? To address this we require a representation which couples 
these two properties together. In 1994 at a dwarf galaxy meeting Bruno 
Binggeli showed a schematic bivariate-brightness distribution (BBD)
for the Virgo cluster. 
In this rather inspirational diagram, galaxies from the Virgo cluster were 
shown on a plot of surface brightness versus luminosity. The Virgo population 
was seen to lie along a luminosity-surface brightness relation with some 
degree of natural morphological segregation (Binggeli 1994). 
Can we construct a similar plot, from existing galaxy catalogues for the 
general field population ?  The concept is 
illustrated in Figure 4 which attempts to convey the usefulness of Binggeli's 
plot to simultaneously reconcile luminosity, surface brightness and morphology.
Figure 4 is divided into three distinct panels the central BBD or Galaxy
Population (GP)-plot,
an upper panel representing the luminosity distribution (obtained from summing
across the surface brightness axis in intervals of luminosity) and a side panel
representing the surface brightness distribution. Currently there exist three 
datasets for which a BBD or GP-plot can be derived, these are: the Two-degree 
Field Galaxy Redshift Survey; the Hubble Deep Fields; and the Local Group. 
These three datasets are shown as Fig. 5(top), 2dFGRS; Fig. 5(middle), HDF;
and Fig 5(bottom), LG and discussed independently below.

\subsection{The Two-degree Field Galaxy Redshift Survey: Cross et al 2000}
The 2dFGRS is an Anglo-Australian collaboration which will eventually obtain
250,000 redshifts for galaxies with $b_{j} < 19.45$ mags selected from UK 
Schmidt plates ($\mu_{lim}=24.67$ mags/$\Box''$). The survey will cover 2000 
sq degrees in two continuous regions (north and south Galactic caps) plus a 
number of randomly located fields. Currently $\geq 100,000$ redshifts have 
been obtained and the expectation is that the survey will be completed by the 
time these proceedings are published. The prime goal of the survey is to study 
large scale clustering and to derive the power spectrum of high surface 
brightness luminous galaxies over large $\sim 100$ Mpc scales.
Figure 5 (upper) shows the range in surface brightness and luminosity
over which 2dFGRS galaxies are seen. Note the data is the observed distribution
uncorrected for the volume bias. The galaxy magnitudes and surface 
brightnesses have been corrected for light lost below the isophote assuming
perfect face-on exponential profiles (see Cross et al 2000 for details). Also
shown is the selection line derived from visibility theory which defines the 
window through which galaxies are seen with volumes $>10^4$Mpc$^3$.

\subsection{The Hubble Deep Field: Driver 1999}
The superb image quality and depth of the HDF provides
an insight into both the distant luminous universe {\it AND} the local 
intrinsically faint universe. Supplementing the spectroscopic redshifts
with photometric redshifts where necessary allows us to extract a 
{\it volume-limited} sample over a well defined range of luminosity and 
surface brightness yielding Figure 5 (middle). The full details of how
this sample and the selection lines were derived are in Driver (1999). 

\subsection{The Local Group: Mateo 1998}
Finally the local group, albeit an ad hoc sample, was recently summarised
by Mateo (1998). While further members of the local group are likely to
be found this sample represents the {\it ONLY} current insight into the 
ultra low luminosity and low surface brightness regimes. Figure 5 (bottom) 
shows the derived GP plot from this sample with approximate selection lines 
shown (these selection lines are based on the depths to which recent surveys
have probed; e.g. Armandroff et al 1999).

\begin{figure}[p]

\vspace{-5.75cm}

\plotone{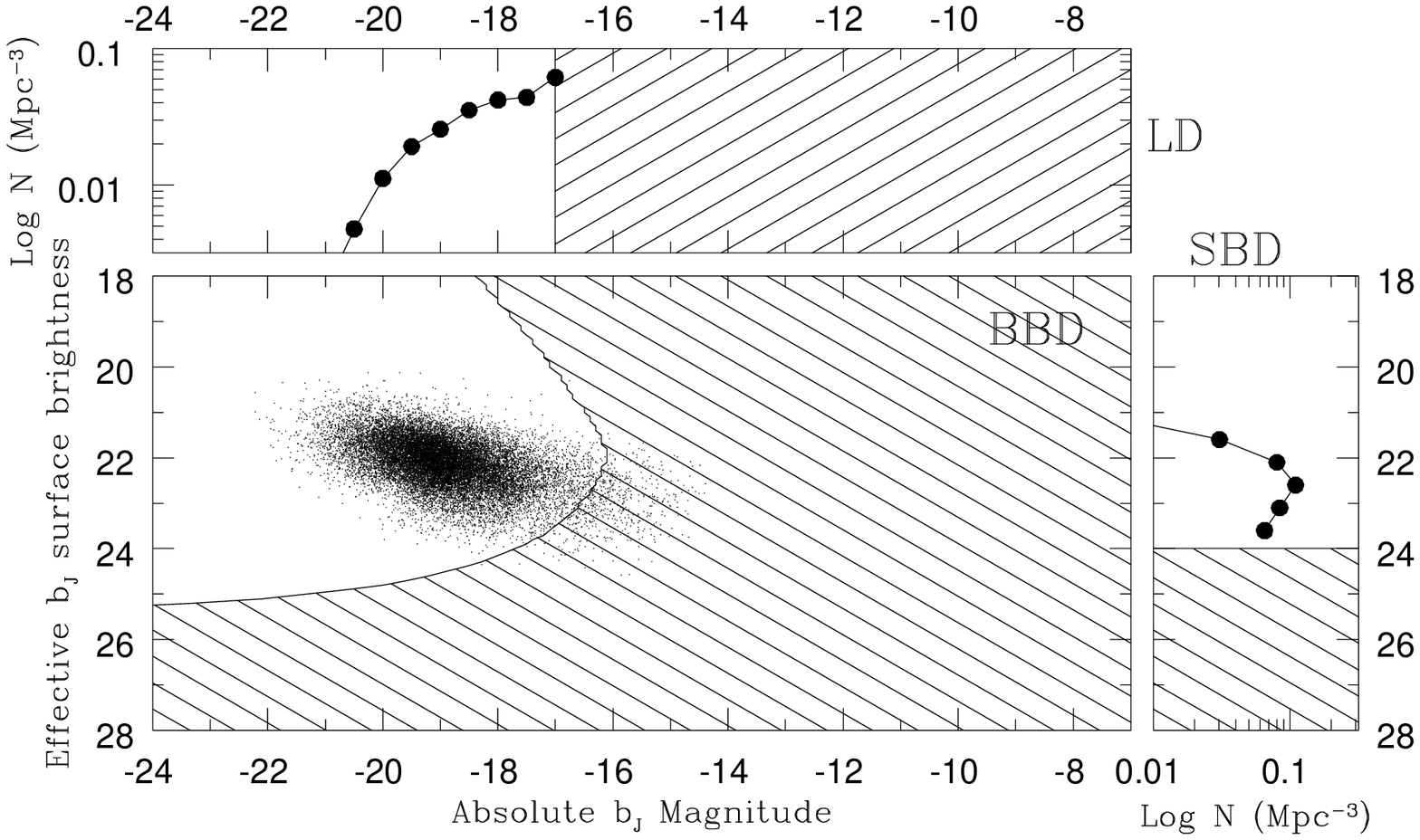}

\vspace{-6.0cm}

\plotone{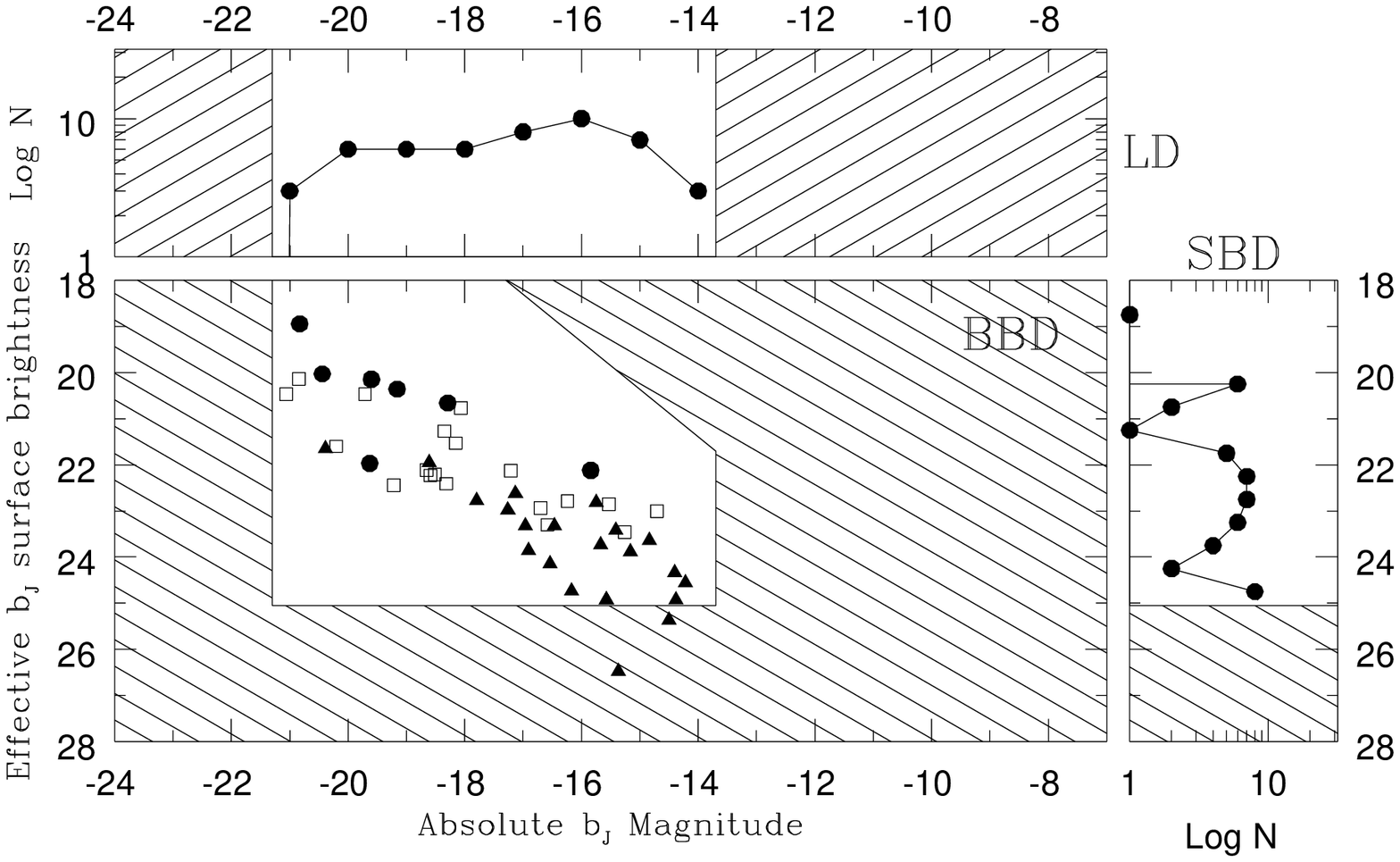}

\vspace{-6.0cm}

\plotone{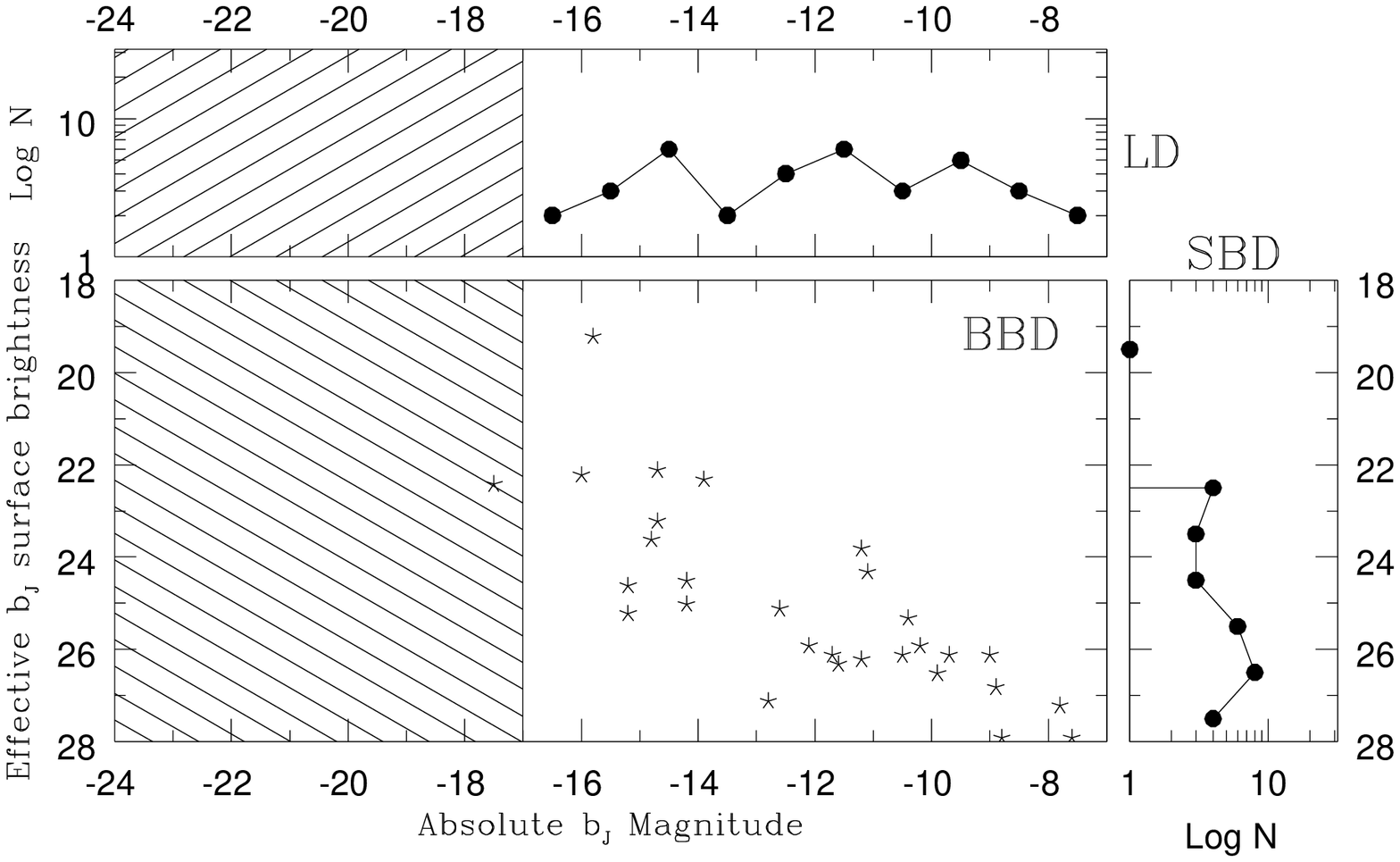}

\vspace{-0.25cm}

\caption{The bivariate brightness distributions recovered from:
(top) the two degree field; (middle) the Hubble deep field; and
(bottom) the local galaxy group}
\end{figure}

\subsection{An $L-\Sigma$ relation and the paucity of
luminous LSBGs}
So what can we derive from these plots ? The main
impression is that each dataset probes a very distinct region of the
possible parameter space over which galaxies can exist. Together they
arguably define the full extent of the low redshift population.

The 2dFGRS sample, while containing the most galaxies ($\sim 20,000$ shown 
here), spans the smallest although possibly the most important region. This 
is because the sample is 
{\it magnitude-limited} and therefore strongly biased towards sampling
high luminosity systems (this is perfectly adequate, if not ideal, for the 
purposes of mapping out large scale structure). The upper and side panels
show the volume-adjusted luminosity and surface brightness distributions 
respectively. We see that the bright end of both distributions are well
defined (as also seen in Figs. 1 and 2.). The 2dFGRS shows a clear 
luminosity-surface brightness trend, indicating that lower surface brightness
systems are of lower luminosity. This may well explain part or all of the 
variation seen in Fig. 1, for example a survey with a bright isophotal limit
will preferentially select against dwarf systems and return a falling faint 
end slope. The 2dFGRS data also shows clear space between
the bulk of the luminous galaxies and the selection boundary. This suggests
that luminous core-less low surface brightness galaxies are either rare or
form an entirely separable population at ultra-low surface brightnesses.

The HDF provides an intermediate window overlapping in parameter space with 
both the 2dFGRS and LG samples and includes morphological information 
(circles, squares and triangles denoting ellipticals, spirals and irregulars
respectively). The HDF and LG data are both 
{\it volume-limited} and the data shown directly reflects the relative
abundance of each galaxy type (within the optical window defined by the
selection limits). The HDF confirms the paucity 
of luminous low surface brightness
galaxies and the trend of the luminosity-surface brightness relation,
albeit with a slightly steeper slope. The LG data shows that this
luminosity-surface brightness relation continues to the faintest limits.

The luminosity and surface brightness distributions of the HDF and LG data
are relatively flat showing that over the previously unexplored range of
$-17 < M_{B} < -8$ and $21 < \mu_{e} < 28$, galaxies exist in equal numbers
up to the selection limits. 

\section{The combined Optical Window}
Figure 6 shows the combination of these three datasets along with
a low surface brightness galaxy sample of de Blok et al (1996), Malin 1 and
a smattering of globular clusters. The densely shaded area illustrates the 
optical parameter space which has {\it NOT} been explored by any survey. The 
lightly shaded area is the region that has {\it ONLY} been explored within
the local group (radius $< 1$ Mpc). This region should be taken with caution
as we are unaware as to how representative the local group may be of the low 
luminosity universe at large - for example the recent results from Virgo
paint a dramatically different, and deeply alarming (divergent), picture of
the low luminosity population (see Phillipps et al 1998).
As before the upper and side panels of Fig. 6 represent the luminosity 
and surface brightness distributions derived from the combination
of our three datasets. These have been crudely derived
by scaling the Hubble Deep Field (covering $\sim 300$ Mpc$^{3}$) and
the local group (covering $\sim \pi$ Mpc$^{3}$ until they overlap with the
faint end of the 2dFGRS data (covering $> 10^{4}$ Mpc$^{3}$).
Both distributions are relatively flat suggesting equal numbers of galaxies
at all luminosity and surface brightness intervals up until some bright
luminosity or surface brightness cutoff. The central plot shows the 
luminosity-surface brightness relation seen in each of the individual datasets.
The low surface brightness galaxy sample of de Blok occupies a fairly unique
region reflecting their unique selection criterion (see Figure 4). Malin 1
lies in the totally unexplored region.
{\bf Fig. 6 represents our combined optical window into the low redshift galaxy
population, it represents the parameter space over which galaxies are known 
to exist coupled within carefully defined selection boundaries.}

\begin{figure}[p]
\plotfiddle{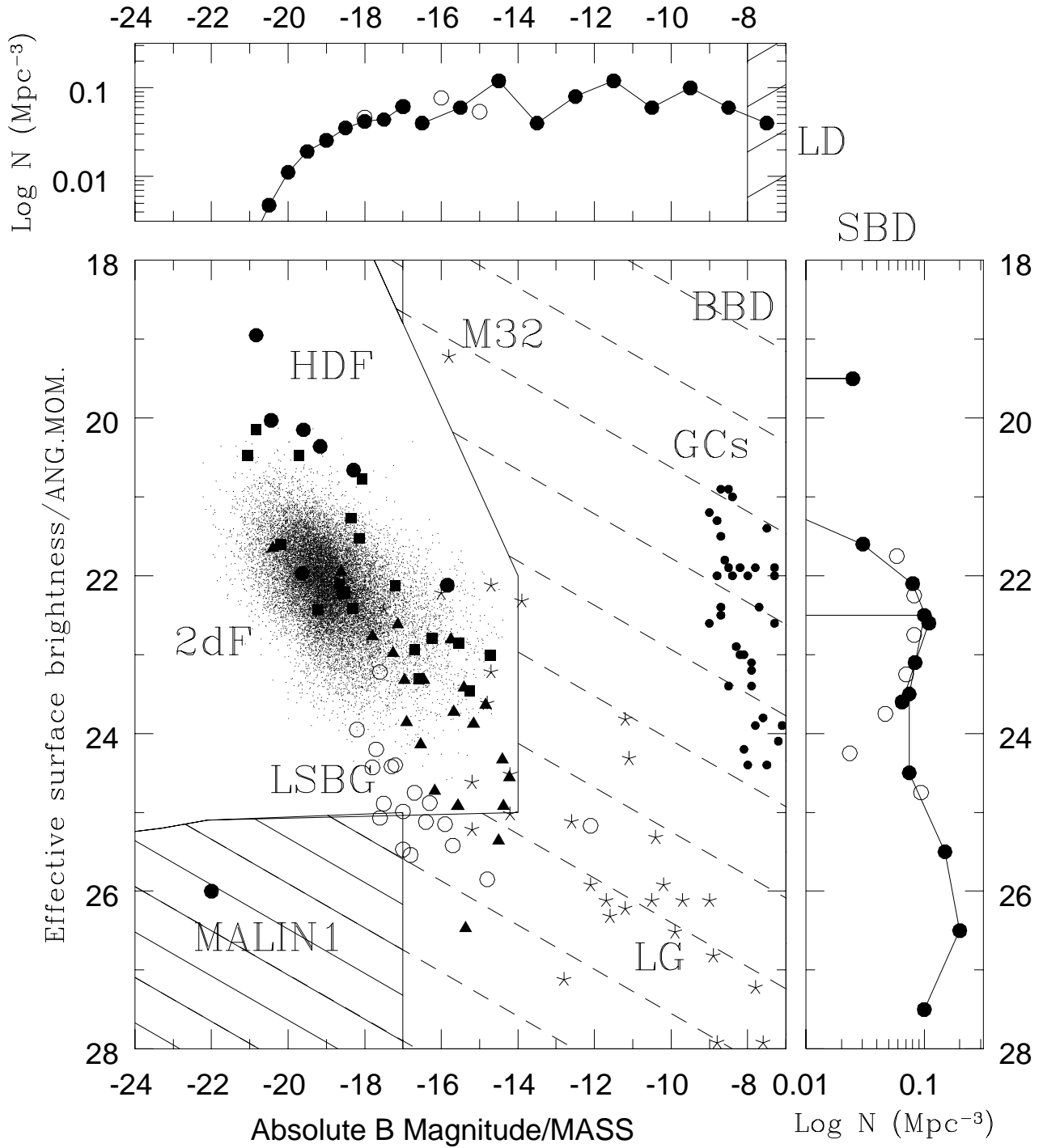}{200mm}{0}{105}{105}{-250}{-150}
\caption{The optical window of galaxy existence. The principle data shown are:
the Two-degree Field Galaxy Redshift Survey (Cross et al 2000), the Hubble 
Deep Field (Driver 1999), the Local Group (Mateo 1998), a low surface 
brightness galaxy sample (de Blok et al 1996) and Malin1 (Bothun et al 198?). 
The heavily shaded region denotes areas in which none of the surveys can 
probe. The lightly shaded region is that surveyed by the local group only.}
\end{figure}

\section{Conclusions}
Here we have summarised the current understanding of the field galaxy
luminosity function and the surface brightness distribution. Because of the
clear luminosity-surface brightness relation we conclude that treating these
key parameters as separable can lead to confusion and error. We advocate a
bivariate brightness distribution which allows us to construct a window into
the galaxy population. Using three distinct independent datasets to explore 
this region we map out the locations in luminosity and surface brightness over 
which galaxies are known to exist. The three datasets considered here are the 
Two-degree Field Galaxy Redshift Survey (spanning the high surface brightness
luminous regime), the Hubble Deep Field (spanning the medium surface brightness
medium luminosity regime), and the Local Group (spanning the low luminosity 
regime). Combining these three surveys reveals a consistent picture in which 
{\it BOTH} the luminosity and surface brightness distribution are flat to the 
detection boundaries.
Across the full optical window over which galaxies are known to exist we 
have only censured a quarter-panel of this parameter space. Much work remains 
to be done before we can confidently state that we have a comprehensive and 
complete understanding of the low redshift universe and all that lies within.

\acknowledgements
We thank the organisers for a very enjoyable conference and our colleagues on
the 2dFGRS and HDF teams for assistance in collating the data shown.

\section*{References}

\begin{description}
\vspace{-0.225cm}
\item{\hspace{0.0cm}} Armandroff, T.E., Jacoby G.H., Davies J.E., 1999, AJ,
118, 1220
\vspace{-0.225cm}
\item{\hspace{0.0cm}} Binggeli B., 1993 in ESO/OHP Workshop on Dwarf 
Galaxies, Eds Meylan G., 
\& Prugniel, P., (Publ: ESO, Garching), 13
\vspace{-0.225cm}
\item{\hspace{0.0cm}} Bothun, G.D., Impey, C.D., Malin, D.F., Mould, J.R. 
1987, AJ, 94, 23
\vspace{-0.225cm}
\item{\hspace{0.0cm}} Carlberg, R.G., Yee, H.K.C., Ellingson, E., Abraham, 
R., Gravel, P., Morris, S. Pritchet, C.J. 1996, ApJ, 462, 32
\vspace{-0.225cm}
\item{\hspace{0.0cm}} Cross, N., et al, 2000, MNRAS, submitted
\vspace{-0.225cm}
\item{\hspace{0.0cm}} Dalcanton, J.J., Spergel, D.N., Gunn, J.E., Schmidt, M., Schneider, D.P., 1997, AJ, 114, 635
\vspace{-0.225cm}
\item{\hspace{0.0cm}} de Blok, W.J.G., McGaugh, S.S., Van der Hulst, J.M. 
1996, MNRAS, 283, 18
\vspace{-0.225cm}
\item{\hspace{0.0cm}} de Lapparent, V., Geller M.J., Huchra J.P., 1986, ApJ, 
304, 585
\vspace{-0.225cm}
\item{\hspace{0.0cm}} Disney, M., 1976, Nature, 263, 573
\vspace{-0.225cm}
\item{\hspace{0.0cm}} Disney, M., \& Phillipps S., 1987, Nature, 329, 203
\vspace{-0.225cm}
\item{\hspace{0.0cm}} Driver, S.P. 1999, ApJL, 526, L69.
\vspace{-0.225cm}
\item{\hspace{0.0cm}} Driver, S.P., et al, 1998, ApJL, 496, L93 
\vspace{-0.225cm}
\item{\hspace{0.0cm}} Efstathiou, G., Ellis, R., Peterson, B. 1988, MNRAS, 
232, 431
\vspace{-0.225cm}
\item{\hspace{0.0cm}} Ellis, R.S., Colless, M., Broadhurst, T., Heyl, J., 
Glazebrook, K. 1996, MNRAS, 280, 235
\vspace{-0.225cm}
\item{\hspace{0.0cm}} Ferguson, H., Binggeli B., 1994, A\&ARv, 6, 67
\vspace{-0.225cm}
\item{\hspace{0.0cm}} Freeman, K. 1970, ApJ, 160, 811
\vspace{-0.225cm}
\item{\hspace{0.0cm}} Fukugita M., Hogan C.J., Peebles P.J.E., 1998, ApJ, 
503, 518
\vspace{-0.225cm}
\item{\hspace{0.0cm}} Impey, C., Bothun, G. 1997, ARA\&A, 35, 267 
\vspace{-0.225cm}
\item{\hspace{0.0cm}} Lin, H., Kirshner, R., Shectman, S., Landy, S., Oemler,
A.,  Tucker, D., Schechter, P. 1996, ApJ, 464, 60
\vspace{-0.225cm}
\item{\hspace{0.0cm}} Loveday, J., Maddox, S. J., Efstathiou, G., Peterson, 
B. A. 1995, ApJ, 442, 457
\vspace{-0.225cm}
\item{\hspace{0.0cm}} Marzke, R., Huchra, J., Geller M. 1994, ApJ, 428, 43
\vspace{-0.225cm}
\item{\hspace{0.0cm}} Marzke, R., Da Costa, N., Pelligrini, P., Willmer, C., 
Geller M. 1998, ApJ, 503, 617
\vspace{-0.225cm}
\item{\hspace{0.0cm}} Mateo, M.L. 1998, ARA\&A, 36, 435
\vspace{-0.225cm}
\item{\hspace{0.0cm}} Naim, A., et al 1995, MNRAS, 274, 1107
\vspace{-0.225cm}
\item{\hspace{0.0cm}} O'Neil, K., Bothun G., 2000, ApJ, 529, 811
\vspace{-0.225cm}
\item{\hspace{0.0cm}} Persic, M., Salucci, P., 1992, MNRAS, 258, 14pp
\vspace{-0.225cm}
\item{\hspace{0.0cm}} Phillipps, S., Davies, J., Disney, M. 1990, MNRAS, 242, 
235
\vspace{-0.225cm}
\item{\hspace{0.0cm}} Phillipps, S., Parker Q.A., Schwartzenberg J.M., 
Jones J.B., 1998, ApJ, 493, 59
\vspace{-0.225cm}
\item{\hspace{0.0cm}} Ratcliffe, A., Shanks, T., Parker, Q., Fong R. 1998, MNRAS, 293, 197
\vspace{-0.225cm}
\item{\hspace{0.0cm}} Schechter, P. 1976, ApJ, 203, 297
\vspace{-0.225cm}
\item{\hspace{0.0cm}} Sprayberry D., Impey, C., Irwin, M., 1996, ApJ, 463, 535
\vspace{-0.225cm}
\item{\hspace{0.0cm}} Zucca, E., et al 1997, A\&A, 326, 477
\vspace{-0.225cm}
\item{\hspace{0.0cm}} Zwaan, M., van der Hulst, J., de Blok, W., McGaugh, S. 
1995, MNRAS, 273 35L
\vspace{-0.225cm}
\item{\hspace{0.0cm}} Zwicky, F., 1957, in Morphological Astronomy,
(Publ: Springer-Verlag)
\end{description}

\end{document}